\documentclass[twocolumn,astrosymb,twocolappendix]{aastex631}

\newcommand{\sbunit}{\mathrm{mag\ arcsec}^{-2}}
\newcommand{\sbcen}{\mu_{0}(g)}
\newcommand{\sbeff}{\overline{\mu}_{\mathrm{eff}}(g)}

\newcommand{\code}[1]{\texttt{#1}}
\newcommand{\sersic}{S\'ersic}
\usepackage{CJKutf8}
\usepackage{bm}
\usepackage{appendix}
\usepackage{amsmath,amssymb}
\usepackage{lipsum}

\defcitealias{Greco2018}{G18}
\defcitealias{Li2022}{L22}

\shorttitle{Quenching of mass--size Outliers}
\shortauthors{Li et al.}
\graphicspath{{./}{figures/}}

\begin{document}
\begin{CJK*}{UTF8}{gbsn}

\title{Beyond Ultra-diffuse Galaxies. II. Environmental Quenching of Mass--Size Outliers among the Satellites of Milky Way Analogs}

\correspondingauthor{Jiaxuan Li}
\author[0000-0001-9592-4190]{Jiaxuan Li (李嘉轩)}
\affiliation{Department of Astrophysical Sciences, 4 Ivy Lane, Princeton University, Princeton, NJ 08544, USA}
\email{jiaxuanl@princeton.edu}
\author[0000-0002-5612-3427]{Jenny E. Greene}
\affiliation{Department of Astrophysical Sciences, 4 Ivy Lane, Princeton University, Princeton, NJ 08544, USA}
\author[0000-0003-4970-2874]{Johnny Greco}
\affiliation{Department of Astrophysical Sciences, 4 Ivy Lane, Princeton University, Princeton, NJ 08544, USA}
\affiliation{Center for Cosmology and AstroParticle Physics (CCAPP), The Ohio State University, Columbus, OH 43210, USA}
\author[0000-0002-1691-8217]{Rachael Beaton}
\affiliation{Department of Astrophysical Sciences, 4 Ivy Lane, Princeton University, Princeton, NJ 08544, USA}
\author[0000-0002-1841-2252]{Shany Danieli}
\altaffiliation{NASA Hubble Fellow}
\affiliation{Department of Astrophysical Sciences, 4 Ivy Lane, Princeton University, Princeton, NJ 08544, USA}
\author[0000-0003-4700-663X]{Andy Goulding}
\affiliation{Department of Astrophysical Sciences, 4 Ivy Lane, Princeton University, Princeton, NJ 08544, USA}
\author[0000-0003-1385-7591]{Song Huang (黄崧)}
\affiliation{Department of Astrophysical Sciences, 4 Ivy Lane, Princeton University, Princeton, NJ 08544, USA}
\affiliation{Department of Astronomy and Tsinghua Center for Astrophysics, Tsinghua University, Beijing 100084, China}
\author[0000-0002-0332-177X]{Erin Kado-Fong}
\affiliation{Department of Astrophysical Sciences, 4 Ivy Lane, Princeton University, Princeton, NJ 08544, USA}
\affiliation{Physics Department, Yale Center for Astronomy \& Astrophysics, PO Box 208120, New Haven, CT 06520, USA}



\begin{abstract}
Recent observations have reignited interest in a population of dwarf galaxies that are large and diffuse for their mass, often called ultra-diffuse galaxies.
However, the origin and evolution of these mass--size outliers and the role of the environment are still unclear. Using the exquisitely deep and wide Hyper Suprime-Cam Strategic Survey images, we search for ultra-puffy galaxies, defined as being $1.5\sigma$ larger than the average size for their mass, around Milky Way-like galaxies. 
We present the sizes and radial distributions of mass--size outliers and derive their quenched fraction to explore the impact of the environment. 
Surprisingly, despite being outliers in size, the ultra-puffy galaxies have a similar quenched fraction as normal-sized satellites of Milky Way analogs in both observations and simulations, suggesting that quenching is not tied to being a mass--size outlier. 
The quenched fraction is higher for the ultra-puffy galaxies associated with redder hosts, as well as those that are closer to the host in projection. 
In contrast, the ultra-diffuse galaxies are overall redder and more quiescent compared with normal satellites. We show that the classic definition of ultra-diffuse galaxies is heavily weighted toward quenched galaxies and thus cannot be used for a study of quenching of mass--size outliers.
\end{abstract}


\section{Introduction} \label{sec:intro}
Dwarf galaxies, given their overwhelming abundance, shallow gravitational potential wells, diverse evolutionary histories, and high dark matter content, are ideal probes for testing the $\Lambda$CDM paradigm and understanding dark matter \citep{Bullock2017} and baryonic processes \citep{Sales2022}. In particular, dwarf galaxies and their star formation history are sensitive to the environments where they reside.  
Nearly all isolated dwarf galaxies in the field are star-forming \citep{Geha2012}, but satellite dwarf galaxies in the Local Group and Local Volume hosted by Milky Way (MW)-mass galaxies are much more quiescent in star formation, with near-unity quenched fraction at $M_\star \lesssim 10^{8}\ M_\odot$ \citep[e.g.,][]{Grcevich2009, Spekkens2014, Wetzel2015, Baxter2021, SAGA-II, Putman2021, CarlstenELVES2022, Greene2022ELVES,Karunakaran2022}. Such a contrast indicates that environment plays an important role in regulating star formation in dwarf galaxies. 

Several physical processes are believed to be responsible for the cessation of star formation (also known as ``quenching'') in dwarf galaxies. Isolated dwarf galaxies can be quenched by bursty stellar feedback \citep[e.g.,][]{ElBadry2018}, reionization (for ultra-faint dwarfs; \citealt{Bullock2000,Benson2002,Somerville2002,Tollerud2018,Applebaum2021}), or previously passing a more massive halo, as in the case of so-called backsplash galaxies \citep{Simpson2018,Benavides2021}. 
For the satellite galaxy of an MW analog (a galaxy having a similar stellar or halo mass as the MW), the ram pressure experienced by the satellite as it moves through the hot circumgalactic medium (CGM) of the host galaxy could remove the gas and cause the cessation of star formation \citep[e.g.,][]{GunnGott1972,McCarthy2008,Grcevich2009,Emerick2016,Simpson2018,Tremmel2020,Samuel2022}. It has been shown that ram pressure stripping can 
efficiently quench galaxies at $M_\star < 10^{8}\ M_\odot$ \citep{Fillingham2016,Simpson2018,Buck2019}. Starvation, where cold gas supplies are halted following infall to a more massive hot halo, is believed to be dominant for higher-mass satellites \citep{Fillingham2016}. Tidal stripping can also remove the gas from the satellite and cause quenching \citep{Simpson2018}, but strong tidal interactions are needed because the dark matter needs to be stripped before a significant gas loss happens. 
A satellite dwarf galaxy might also be preprocessed in a lower-mass group prior to falling into the current host galaxy, and such preprocessing would increase the quenched fraction \citep{Wetzel2015b,Jahn2022,Samuel2022}. Extensive studies have been done on the quenching of satellites in MW analogs using simulations \citep[e.g.,][]{Simpson2018,Buck2019,Garrison-Kimmel2019,Simons2020,Akins2021,Joshi2021,Karunakaran2021,Font2022,Pan2022,Samuel2022}, and many of them are able to reproduce the observed quenched fraction in the Local Group and Local Volume. It is now time to understand the relative importance of these mechanisms as a function of satellite and host properties.

Interestingly, several of these physical processes might be able to puff up satellite galaxies and turn them into outliers with respect to the average mass--size relation. An example class of mass--size outlier is the ultra-diffuse galaxies (UDGs, \citealt{vanDokkum2015}), which have a large size ($r_e > 1.5$~kpc) and low surface brightness ($\sbcen > 24\ \sbunit$). 
It is therefore valuable to ask if the same mechanism that dominates quenching also produces the observed population of mass--size outliers. 
This will provide a unique angle to peek into the formation and evolution of satellite galaxies in MW analogs and help us understand whether the mass--size outliers are a distinct class of satellite galaxies or just the large-size tail of the satellite population. 

Large samples of mass--size outliers already exist in galaxy cluster environments, but lower-density environments such as groups are relatively unexplored \citep[e.g.,][]{Karunakaran2022b}. In \citet[][hereafter \citetalias{Li2022}]{Li2022}, we presented a sample of dwarf galaxies that are mass--size outliers among the satellites of MW analogs at $0.01 < z < 0.04$. As a follow-up study, in this paper, we ask whether the mass--size outliers are more quiescent compared to the ``normal'' satellite population of MW analogs. 
We take the mass--size outliers from \citetalias{Li2022} (described in \S\ref{sec:data}) and present their characteristics, including their size and radial distributions, in \S\ref{sec:sample_char}. We then derive the quenched fractions as a function of satellite stellar mass, host color, and projected distance to the host (\S\ref{sec:quench}). We compare the quenched fractions of mass--size outliers with the bulk of the satellite population of nearby MW analogs and simulations. In \S\ref{sec:discussion}, we discuss the implications of these results on the formation and quenching of mass--size outliers. We adopt a flat $\Lambda$CDM cosmology from \citet{Planck15} with $\Omega_{\rm m}= 0.307$ and $H_0 = 67.7\ $km s$^{-1}$ Mpc$^{-1}$. We use the AB system \citep{Oke1983} for magnitudes. The stellar masses used in this work are based on the \citet{Chabrier2003} initial mass function.


\section{Data and Sample Selection} \label{sec:data}
In this section, we briefly summarize the data and the mass--size outlier samples presented in \citetalias{Li2022}. We refer interested readers to \citetalias{Li2022} for a detailed description of the samples and the search algorithms. 

\subsection{mass--size outlier sample}
In \citetalias{Li2022}, we performed a systematic search for low surface brightness galaxies (LSBGs) in Subaru's Hyper Suprime-Cam Strategic Survey Program (HSC-SSP, \citealt{Aihara2018}; hereafter the HSC survey). The HSC survey is an imaging survey using the 8.2-m Subaru telescope and the Hyper Suprime-Camera \citep{Miyazaki2012, Miyazaki2018}, covering $\sim 1000\ \rm{deg}^2$ in five broad bands ($grizy$) and reaching a depth of $g=26.6$ mag, $r=26.2$ mag and $i=26.2$ mag ($5\sigma$ point source detection). In \citetalias{Li2022}, we used the data release PDR2, which covers $\sim 300\ \rm{deg}^2$ and has a global sky subtraction to preserve low surface brightness features \citep{Aihara2018,Li2021}. 

We search for LSBGs following the method in \citet{Greco2018} with several modifications to suit PDR2 and improve the completeness and purity. Briefly, we run \code{SExtractor} \citep{Bertin1996} on the coadd images after removing bright extended sources and associated diffuse light. After applying an initial size and color cut to the output catalog, we model each source by running \code{scarlet} \citep{Melchior2018} and extract a series of structural and morphological parameters from the nonparametric model. We design a metric based on the color, size ($r_e$, circularized half-light radius), surface brightness ($\sbeff$, average surface brightness within $r_e$ in the $g$ band), and morphology and use it to remove false positives that are not likely to be real LSBGs. For the remaining objects, we fit a parametric model to estimate their sizes, magnitudes, and surface brightnesses. The completeness of the search is characterized by injecting mock galaxies into coadd images and recovering them with our algorithm. In the size range of $3\arcsec < r_e < 14\arcsec$, our search is $>70\%$ complete to $\sbeff < 26.5\ \sbunit$ and $>50\%$ complete to $\sbeff \leqslant 27.0\ \sbunit$. For comparison, \citet{Zaritsky2021} searched for LSBGs in the Dark Energy Camera Legacy Survey and reached $\sim25\%$ completeness at $\mu_0(g)\approx 25.5\ \sbunit$. \citet{Tanoglidis2021} reported the LSBG search using Dark Energy Survey data and showed a completeness $\sim 30\%$ at $\sbeff \approx 26\ \sbunit$. Therefore, our completeness is high compared with other LSBG searches, which ensures high fidelity in the scientific results. We remove LSBGs with a completeness of less than 10\% for cleaner statistics. 

Then we cross-match our LSBG sample with MW analogs selected from the NASA-Sloan Atlas\footnote{\url{http://nsatlas.org}} \citep{Blanton2005,Blanton2011} catalog by requiring $0.01 < z < 0.04$, $10.2 < \log\ M_\star/M_\odot < 11.2$, and being in the HSC PDR2 footprint. In this way, we select 922 MW analogs. For context, we compare our selection with two surveys that also search satellites of MW analogs. The Satellites Around Galactic Analogs \citep[SAGA;][]{SAGA-I,SAGA-II} survey is an ongoing spectroscopic survey of classical satellites of 100 MW analogs at $20$~Mpc~$< D < 40$~Mpc. Their MW analogs are in a stellar mass range of $10.2 < \log\ M_\star/M_\odot < 11.0$. The Exploration of Local VolumE Satellites \citep[ELVES;][]{Carlsten2020Radial,ELVES-I,ELVES-II,CarlstenELVES2022} survey maps the satellites of 30 host galaxies with $M_\star > 10^{9.9}\ M_\odot$ and $D<12$~Mpc. Most of the satellites have direct distances from either the tip of the red giant branch or surface brightness fluctuations, with the latter dominating for the faint satellites. The ELVES sample already includes MW and M31 satellites. To sum up, our MW analog selection is similar to that of SAGA. Because of the volume-limited criterion, ELVES includes several groups more massive than the SAGA hosts and ours. 

We calculate the virial mass of MW analogs based on the stellar-to-halo mass relation in \citet{Behroozi2010} using \href{https://halotools.readthedocs.io/en/latest/index.html}{\code{halotools}} \citep{Hearin2017}. Then we convert virial mass to virial radius $R_{\rm vir}$ assuming a flat $\Lambda$CDM cosmology where the virial overdensity only depends on cosmological parameters and redshift \citep{Bryan1998}. For a given MW analog, we associate any LSBG that falls inside its projected virial radius with the MW analog as the host galaxy. If one LSBG is matched to more than one host, we assign it to the nearest host based on the angular separation normalized by the host virial radius. In total, we surveyed the 922 MW analogs, and we found 2510 LSBG candidates associated with 689 MW analogs. We assume that the LSBGs are at the same redshifts as their hosts. 

After cross-matching, we select two mass--size outlier samples, namely UDGs and ultra-puffy galaxies (UPGs). The UDGs are defined to have $r_e + \sigma(r_e) > 1.5\ \mathrm{kpc}$ and $\sbeff + \sigma(\sbeff) > 25\ \sbunit$, which takes the $1\sigma$ measurement error into account. This definition is consistent with the original definition in \citet{vanDokkum2015} for galaxies with \sersic{} indices $n\sim 1$. There are 412 LSBG candidates that satisfy the UDG definition, and they are defined as our UDG sample. As we argued in \citetalias{Li2022}, the UDG definition does not consider the dependence of galaxy size on galaxy mass. A UDG with $r_e = 1.5$~kpc is an outlier in size if its stellar mass is $10^7\ M_\odot$, but it will be normal-sized if its stellar mass is $10^{8.5}\ M_\odot$. Therefore, UDGs are not necessarily mass--size outliers. We propose the concept of \textit{ultra-puffy galaxies}, which are defined to lie $1.5\sigma$ above the average mass--size relation of the satellite galaxies. Because we are interested in mass--size outliers associated with MW analogs, we take the mass--size relation and its scatter from \citet{ELVES-I},  which is derived from the satellites of MW analogs in the Local Volume. The mass--size relation in \citet{ELVES-I} is measured for $10^{5.5} < M_\star / M_\odot < 10^{8.5}$, and we linearly extrapolate this relation to $M_\star \sim 10^{9}\ M_\odot$ to select UPGs in this work. Among our LSBG candidates, we have 337 galaxies as our sample of UPGs that fall $1.5\sigma$ above the average mass--size relation. The UPGs are associated with 239 MW analogs. 

\subsection{Contamination correction}\label{sec:bkg}

We note that the LSBGs are associated with the MW-like hosts only in projection, so a certain fraction of them will be foreground or background galaxies that happen to be close to the MW analogs in the sky. Therefore, we apply a statistical interloper correction as follows. We randomly select a continuous patch of the sky of 24~deg$^2$ in HSC PDR2 regardless of whether it contains MW analogs. Then we repeat the LSBG search in this area and apply the same cuts as we described above to remove false positives. There are 480 LSBG candidates in this area that represent possible contaminants for the UDG and UPG samples. Because distances are needed to define the UDG and UPG classes, we randomly match these 480 LSBGs with the 922 MW analogs that we have surveyed. With such ``artificial'' associations, we calculate the corresponding physical sizes and stellar masses and select UDGs and UPGs accordingly. Such matching was repeated 200 times, and we obtain 7625 artificial UDGs and 8267 artificial UPGs. These UDGs and UPGs are ``artificial'' only in the sense of being associated with random MW-like hosts. Therefore, the number densities of artificial UDGs and UPGs are $S_{\rm UDG} = 1.60\pm0.25\ \mathrm{deg}^{-2}$ and $S_{\rm UPG} = 1.72\pm0.23\ \mathrm{deg}^{-2}$, respectively. Equivalently, the contamination fractions for both samples are $f_{\rm contam}^{\rm UDG} \approx 35\%\pm5\%$ and $f_{\rm contam}^{\rm UPG} \approx 45\%\pm6\%$. We have checked that these fractions are not sensitive to the specific sky regions used to construct the artificial UDG and UPG samples \citepalias{Li2022}. We will use these values to correct the contribution from the foreground and background interlopers when deriving the size and radial distributions in \S\ref{sec:size_distr} and \S\ref{sec:radial_distr}.

Furthermore, the contaminants might not be homogeneous in color and could bias the color distribution, thereby changing the quenched fraction of the mass--size outliers. If the artificial UDGs (UPGs) have different properties from the real ones, we can use this information to derive a probability that an object is a real UDG (UPG) at the proposed distance. Indeed, we find that the $g-i$ color distribution of the ``artificial'' UDG (UPG) sample is bluer than that of the observed UDG (UPG) sample. Specifically, the 25th, 50th, and 75th quantiles of the $g-i$ color are (0.50, 0.71, 0.82) for the real UPG sample and are (0.42, 0.54, 0.72) for the artificial UPG sample. The color distribution of the observed UDGs (UPGs) will thus be biased to be bluer probably because interlopers such as field dwarfs and spiral LSBGs tend to be blue. This motivates us to assign importance weights based on the $g-i$ color distribution and the contamination fraction $f_{\rm contam}$. 

Taking UDGs as an example, we first compute the normalized histograms of $g-i$ colors for both observed UDG and artificial UDG samples (denoted as $\lambda_k^{\rm obs}$ and $\lambda_k^{\rm artificial}$, at the $(g-i)_{k}$ bin, where $\sum_k \lambda_k = 1$). The weight assigned to UDGs in color bin $(g-i)_k$ is then estimated to be $w_k = \max\,(1 - f_{\rm contam}^{\rm UDG} \lambda_k^{\rm artificial} / \lambda_k^{\rm obs}, 0)$. This weight stands for the possibility of not being a contaminant. The weights of UPGs are assigned following the same procedure. Such weights are applied when we calculate the quenched fraction in \S\ref{sec:quench}. We test this method by generating a population of mock UDGs (UPGs), adding a bluer population of contaminants, and recovering the underlying UDG (UPG) color distribution and quenched fraction. Such a mock test verifies that this color-based contamination subtraction is sufficient for this work, and our main results are robust against contamination subtraction.

\section{Sample Characteristics}\label{sec:sample_char}
In this section, we present statistical analyses of the mass--size outliers to gain a comprehensive understanding of the sample. We derive the size (\S\ref{sec:size_distr}) and radial (\S\ref{sec:radial_distr}) distributions, paving the way for the discussion of quenching and environmental effects. 

\begin{figure*}
	\vbox{ 
		\centering
		\includegraphics[width=1\linewidth]{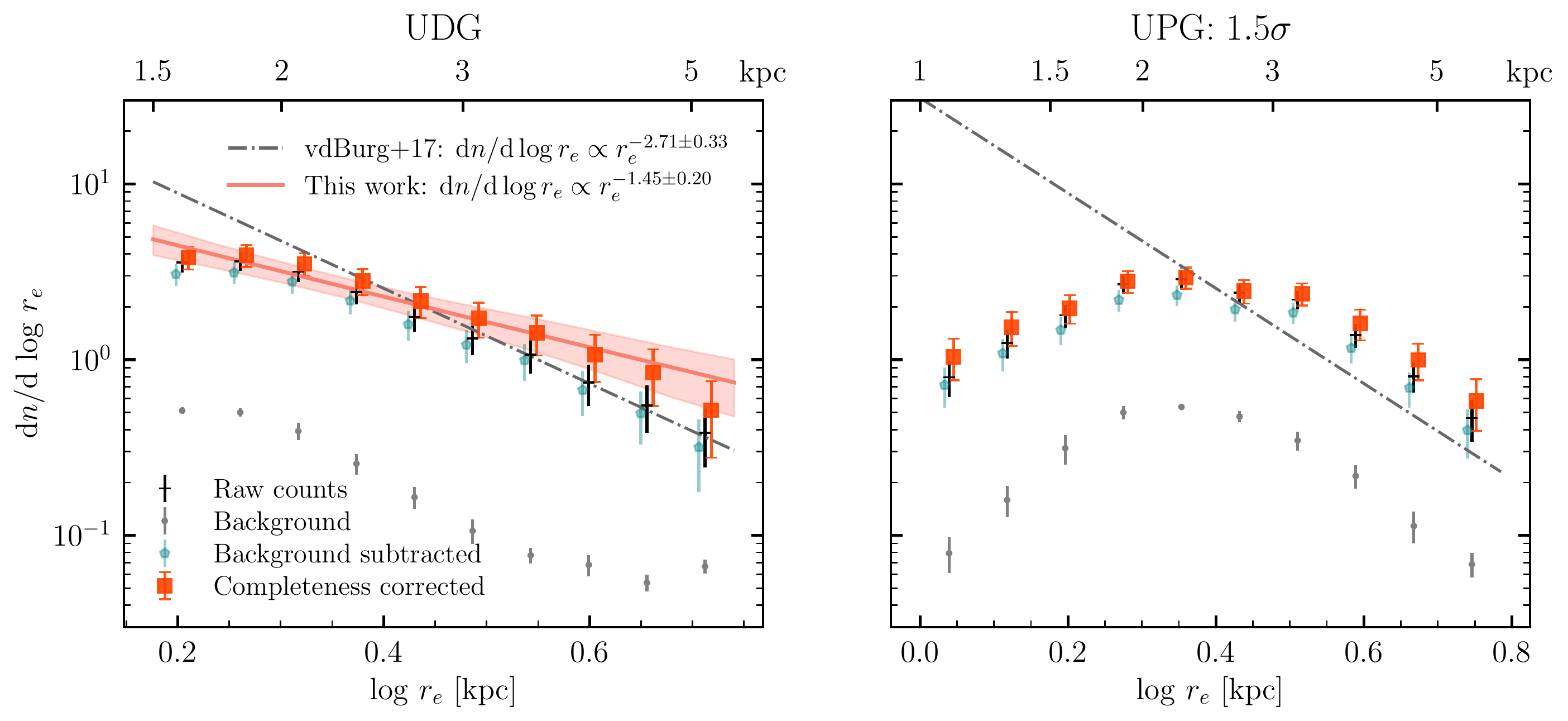}
	}
    \caption{Size distribution of UDGs (\textit{left}) and UPGs (\textit{right}) hosted by MW analogs. The contribution from the foreground and background contaminants (gray dots) is subtracted, and the completeness effect has been corrected. We fit a power law to the size distribution of UDGs and get a power law $\mathrm{d} n / \mathrm{d} \log r_e \propto r_e^{-1.45 \pm 0.20}$. By comparing with the size distributions of UDGs in \citet{vdBurg2016,vdBurg2017}, we find that the power law becomes steeper as the average halo mass increases. It is possible that denser environments provide stronger tidal forces and thus produce more small UDGs. The UPG size distribution is suppressed at the smaller-size end because small galaxies are not $1.5\sigma$ above the mass--size relation and thereby not included as UPGs. 
    }
    \label{fig:size_distribution}
\end{figure*}

\subsection{Size distribution}\label{sec:size_distr}

The size distribution of UDGs is an important topic, since it has been used to test the formation scenarios of UDGs \citep[e.g.,][]{Amorisco2016,vdBurg2017}. In this section, we calculate the size distributions of our UDG and UPG samples and compare them with the literature. We bin the sizes on a logarithmic scale ($\log\, r_e$) and calculate the size distribution $\mathrm{d} n / \mathrm{d} \log r_e$ as follows. First of all, we take the artificial UDGs in the random fields (described in \S\ref{sec:data}) and calculate their size distribution per square degree as a proxy for contamination. For each host, we calculate the size distribution of the associated UDGs, and we multiply the contaminant size distribution by the virial area of the host and subtract it from the UDG size distribution. We combine the UDG size distributions for all hosts by taking the average size distribution, shown as green pentagons in Figure \ref{fig:size_distribution}. For reference, we plot the contaminant size distribution in an average virial area as gray dots. The completeness is calculated for each size bin, and the correction is applied. The final size distribution is shown as red squares. Although the number of mass--size outliers scales with stellar mass, we do not weight each host according to stellar mass because the stellar masses of the hosts in our sample are quite similar. The error shown in Figure \ref{fig:size_distribution} takes both the Poisson error and the measurement error in size into account.

We find that the UDG size distribution roughly follows a power law. We then fit a power law to our UDG size distribution and get $\mathrm{d} n / \mathrm{d} \log r_e \propto r_e^{-1.45 \pm 0.20}$. This is shallower than the power laws presented in \citet[][$\mathrm{d} n / \mathrm{d} \log r_e \propto r_e^{-3.40\pm0.19}$]{vdBurg2016} and \citet[][$\mathrm{d} n / \mathrm{d} \log r_e \propto r_e^{-2.71\pm0.33}$, shown as gray dashed-dotted lines in Figure \ref{fig:size_distribution}]{vdBurg2017}, where they also corrected for background contribution and completeness. Among others, \citet{vdBurg2016} focused on large galaxy clusters and find a quite steep power law; \citet{vdBurg2017} included less-massive groups and the resulting power law is less steep. Combined with our findings, it might be possible that a denser environment produces more small UDGs, probably due to stronger tidal interactions. 

The distribution of $\log\,r_e$ for UPGs is shown in the right panel of Figure \ref{fig:size_distribution}. The size distribution is no longer monotonic because the UPG sample is constructed not by cutting at a certain size but by cutting along the mass--size relation. Consequently, for a given mass, many small UDGs are not in the UPG sample, since their sizes are not extreme on the mass--size plane. The size distribution of the UPG sample is thus suppressed at the smaller size end. We also split the UDG and UPG samples based on their host stellar masses and distance to hosts. For UDGs, we consistently find shallow power laws with indices of $\sim -1.5$, not depending on host stellar mass. Interestingly, we do not find a significant change in the size distribution as UDGs (UPGs) get closer to their hosts.

\subsection{Radial distribution}\label{sec:radial_distr}

\begin{figure*}
	\vbox{ 
		\centering
		\includegraphics[width=1\linewidth]{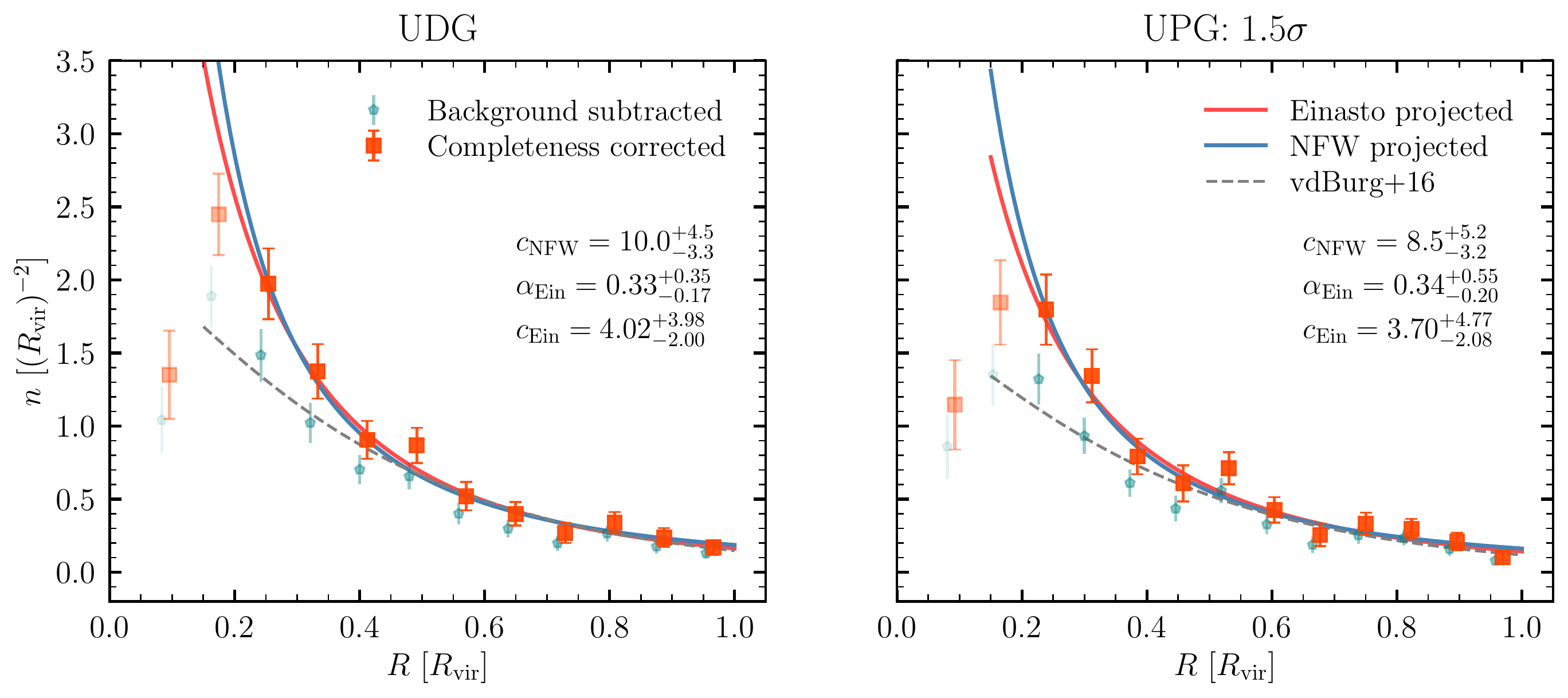}
	}
    \caption{Radial distributions of UDGs (\textit{left}) and UPGs (\textit{right}). We show the radial distributions after subtracting the foreground and background contamination (shown in teal) and correcting for completeness in each radial distance bin (shown in red). After excluding $R<0.2\,R_{\rm vir}$ (shown in lighter colors), we fit projected NFW and Einasto profiles to the radial distributions and find good fits for both profiles. The radial distributions of UDGs and UPGs are similar in shape. The best-fit NFW concentration is consistent with an MW-mass halo, whereas the best-fit Einasto profiles have a smaller concentration than MW-mass halos.}
    \label{fig:radial_distribution}
\end{figure*}

The spatial distribution of satellite galaxies around their host is proposed as a good tracer of the host dark matter halo profile and thus is a probe of environmental effects. Many studies have found that the radial distribution of satellites can be well fit by a Navarro-Frenk-White \citep[NFW;][]{NFW1997} profile, although the radial distribution of satellites is a function of host mass, satellite mass, and satellite color \citep[e.g.,][]{vandenBosch2005,Sales2007,Budzynski2012,Tal2012,Wang2014,Carlsten2020Radial,McDonough2022}. On the other hand, the radial distribution of satellites is a sensitive probe of many physical processes, including reionization \citep{Kravtsov2004} and tidal disruption \citep{Samuel2020}. Subhalos can also be artificially disrupted due to numerical issues in simulations \citep{vandenBosch2018,Carlsten2020Radial}, thus changing the radial distribution of satellites. Furthermore, it is unclear whether the mass--size outliers follow the same radial distribution as normal satellites \citep[e.g.,][]{Tremmel2020}. In cluster environments, the number of UDGs is found to be depleted near the centers of clusters, indicating that they are more likely to be disrupted near the cluster center \citep[e.g.,][]{vanDokkum2015,vdBurg2016,ManceraPina2018}. The radial distribution of mass--size outliers in groups is not as well explored. In this subsection, we derive the radial distributions of the mass--size outliers, compare them with literature results, and discuss possible implications.

We calculate the radial distribution as follows. For each host, we count the number of UDGs (UPGs) in each radial bin scaled by $R_{\rm vir}$ and calculate the number density of UDGs (UPGs) within each radial annulus. Since foreground and background contaminants are distributed quite homogeneously in the sky, their contribution to the observed radial distribution can be subtracted based on the average contaminant density and the angular area occupied by the host. Then we average over all hosts and correct for completeness. The final radial distributions are shown in Figure \ref{fig:radial_distribution}. Red squares are the number of UDGs (UPGs) per radial bin, and the error includes the Poisson, contamination subtraction, and completeness errors. The radial distributions of both UDGs and UPGs turn over at $R<0.2\,R_{\rm vir}$ (shown in lighter colors). Completeness may well drop with proximity to the host due to the blending between the UDG and the host galaxy or due to sky subtraction issues. Although we have corrected each object for completeness based on size and surface brightness, it is still possible that we have not fully corrected for radial incompleteness within a halo. As we mentioned above, it is also possible that UDGs and UPGs are depleted because of tidal disruption. 

We fit the radial distributions of UDGs (UPGs) with the projected NFW and Einasto \citep{Einasto1965} profiles for $R > 0.2 R_{\rm vir}$ using \href{https://bdiemer.bitbucket.io/colossus/index.html}{\code{colossus}} \citep{Diemer2018}. The Einasto profile introduces an extra ``shape'' parameter $\alpha_{\rm Ein}$ and has been argued to be a better description of dark matter halo profiles in simulations \citep[e.g.,][]{Navarro2004,Gao2008,Navarro2010,Dutton2014}. The Einasto profile also makes concentration estimates less sensitive to the radial range fitted. The best-fit parameters are obtained using the least-squares method, and the parameter uncertainty is from the estimated covariance matrix of the parameters. The best-fit NFW (blue) and Einasto (red) profiles are shown in Figure \ref{fig:radial_distribution} as solid lines. The best-fit Einasto profile from \citet{vdBurg2016} is shown as the dashed gray line.

Compared with \citet{vdBurg2016}, there are more UDGs at smaller radial distances in our sample; thus, the UDG radial distribution is more concentrated, even though we exclude $R < 0.2\ R_{\rm vir}$. \citet{vdBurg2016} found that the radial distribution of UDGs cannot be well described by an NFW profile, but an Einasto profile does provide a good fit to the data. Unlike \citet{vdBurg2016}, we find that both the NFW and Einasto profiles describe the radial distribution of UDGs (UPGs) quite well. The concentration of the best-fit NFW profile is $c_{\rm NFW, UDG} = 10.0^{+4.5}_{-3.3}$ for UDGs and $c_{\rm NFW, UPG} = 8.5^{+5.2}_{-3.2}$ for UPGs. The best-fit Einasto profile has $\alpha_{\rm Ein, UDG} = 0.33^{+0.35}_{-0.17},\ c_{\rm Ein, UDG} = 4.02^{+3.98}_{-2.00}$ for UDGs and $\alpha_{\rm Ein, UPG} = 0.34^{+0.55}_{-0.20},\ c_{\rm Ein, UPG} = 3.70^{+4.77}_{-2.08}$ for UPGs. The radial distribution profiles of UDGs and UPGs are found to be very similar in shape. Our best-fit Einasto profiles have lower $\alpha$ and higher concentration compared with \citet{vdBurg2016}, where they have $\alpha_{\rm Ein, UDG} = 0.92^{+0.08}_{-0.18},\ c_{\rm Ein, UDG} = 1.83^{+0.13}_{-0.12}$.

According to the concentration--mass relation of dark matter halos, the concentration of an MW-like halo is $c_{\rm NFW} \sim 10$ \citep[e.g.,][]{Bullock2001,Duffy2008,Dutton2014,Diemer2019}. The Einasto shape parameter for an MW-like halo is $\alpha_{\rm Ein} \sim 0.15$ at $z=0$ and its concentration is $c_{\rm Ein} \sim 8-10$ \citep{Gao2008,Dutton2014}. From Figure \ref{fig:radial_distribution}, the best-fit NFW profiles for the radial distributions agree with the MW-like halo in terms of concentration, but the best-fit Einasto profiles have lower concentration compared with MW-like halos. This agrees with the fact that the concentration of the satellite radial distribution is lower than that of the halo \citep[e.g.,][]{McDonough2022}. We notice that the errors of the best-fit parameters are large due to the sample size and also the fact that concentration is most sensitive to data at smaller radial distances, which are excluded from the analysis. A larger, cleaner, and more complete sample is needed to better constrain the radial distributions. 


\begin{figure*}[htbp!]
	\vbox{ 
		\centering
		\includegraphics[width=1\linewidth]{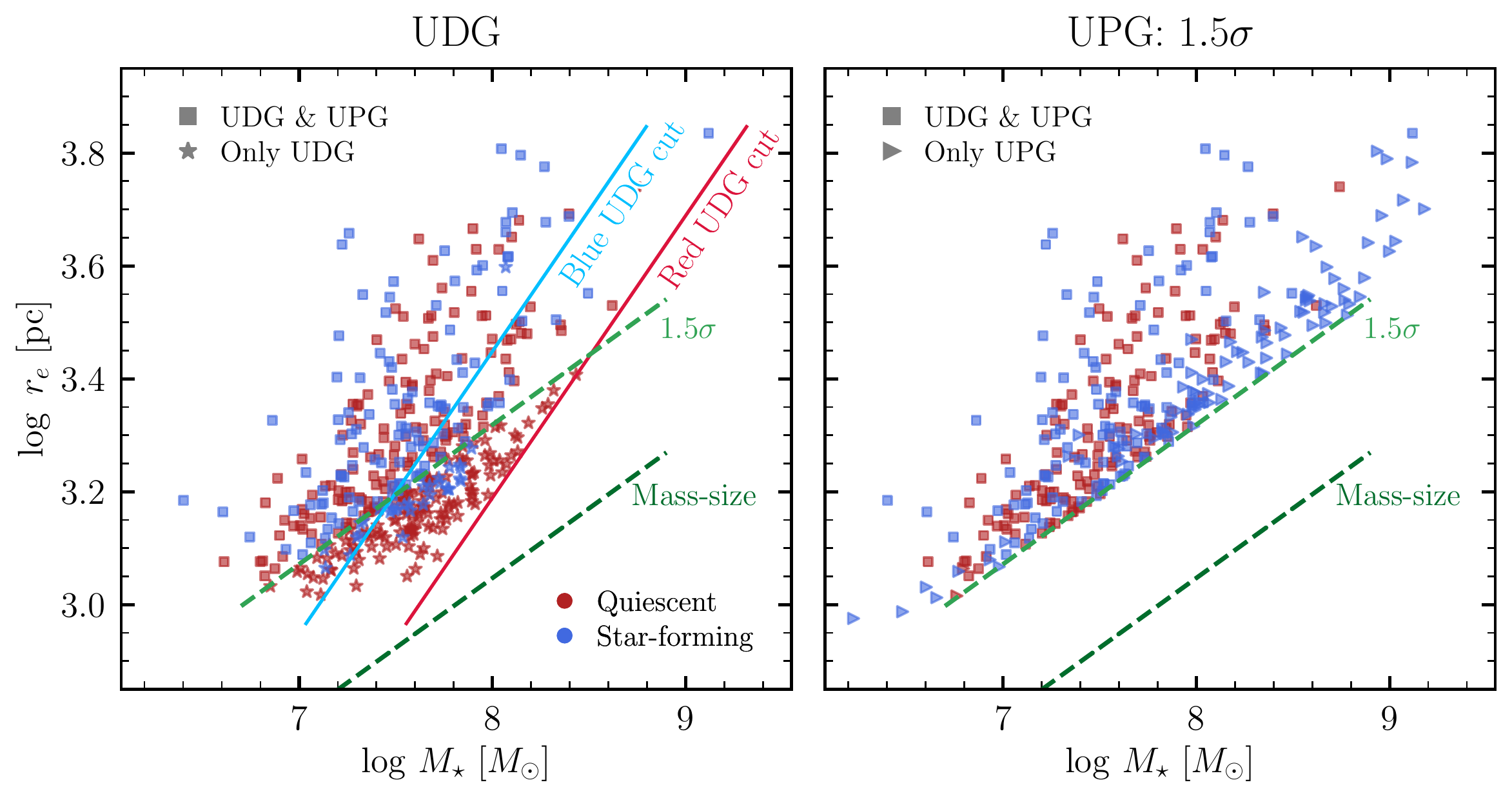}
	}
    \caption{Distribution of UDGs (left) and UPGs (right) on the mass--size plane. We classify galaxies into quiescent and star-forming based on a mass-dependent color cut. The average mass--size relation and $1.5\sigma$ above it are shown in green. The solid lines are constant surface brightness cuts at $\sbeff=25\ \sbunit$ for two different colors $g-i=0.4$ (blue) and $g-i=0.8$ (red). The UDG sample includes a significant number of galaxies that are not mass--size outliers (falling below the $1.5\sigma$ line) that are red in color because the surface brightness cuts are different on the mass--size plane for blue and red galaxies. 
    }
    \label{fig:mass_size}
\end{figure*}

\section{Quenching}\label{sec:quench}
As argued in \S\ref{sec:intro}, the star-forming properties of mass--size outliers are an interesting subject given their diffuse stellar components and unknown formation history. In this section, we first discuss how to define ``quenching'' for our sample and present the distribution of mass--size outliers on the mass--size plane (\S\ref{sec:color}). Then we study the quiescent fractions for UDGs and UPGs as a function of stellar mass and host properties and compare them with other observations and simulations (\S\ref{sec:qfrac}). We discuss how these results would imply the formation and quenching of mass--size outliers in \S\ref{sec:discussion}.

\subsection{Quenching criteria}\label{sec:color}

In principle, direct estimates of the current star formation rate (SFR) from UV or H$\alpha$ or estimates for the \ion{H}{1} gas are needed to define whether a satellite is quenched. For example, the SAGA survey \citep{SAGA-I,SAGA-II} measures the H$\alpha$ equivalent width (EW) from spectra to define quenched satellites. There are also many studies trying to probe \ion{H}{1} for star-forming satellites in the Local Group \citep[e.g.,][]{Grcevich2009,Spekkens2014,Putman2021,Karunakaran2022,Zhu2023} and derive the quiescent fraction. However, due to the low surface brightness and low luminosity nature of our mass--size outliers, obtaining such measurements is extremely expensive over the full footprint of our sample. Therefore, we seek other indicators for SFR, including the broadband colors and morphology. 

The ELVES survey \citep{CarlstenELVES2022} 
searches for satellite galaxies of 30 MW-like hosts in the Local Volume ($D<12$~Mpc). \citet{CarlstenELVES2022} visually inspected all satellites in the ELVES survey and classified them into early-type (red and smooth) and late-type (blue, asymmetric, clumpy). Using this sample, they found that a mass-dependent color cut $(g-i)_{Q} = -0.067 \cdot M_V - 0.23$ could divide the sample into two subsets that are nearly identical to the early and late types based on morphology. \citet{CarlstenELVES2022} derived the quenched fraction of the ELVES satellites using such a color cut and showed that the resulting quenched fractions are very similar to the morphology-based quenched fractions. Furthermore, \citet{Font2022} showed that this color cut effectively separates star-forming galaxies from quiescent ones in simulations that are not fine-tuned to match the observed properties of ELVES galaxies. Therefore, we use this color cut to define quenched galaxies among our mass--size outliers: galaxies that are redder than $(g-i)_Q$ are defined as quenched, whereas bluer galaxies are defined as star-forming.
To apply the color cut, we derive the $V$-band apparent magnitude following $V = g - 0.5784 \cdot (g - r) - 0.0038$ \footnote{This relation was derived for SDSS filters. We neglect the small difference between SDSS and HSC filter systems. \url{http://classic.sdss.org/dr4/algorithms/sdssUBVRITransform.html\#Lupton2005}} and convert it to absolute magnitude $M_V$. For comparison with ELVES and SAGA (Figures \ref{fig:qfrac}, \ref{fig:qfrac_bins}, \ref{fig:qfrac_elves_sim}), we use the same color cut to maintain consistency. 

Before calculating the quenched fraction, we first look at the distributions of mass--size outliers on the mass--size plane, as shown in Figure \ref{fig:mass_size}. This figure is similar to Fig. 5 in \citetalias{Li2022} but here we highlight the quiescent galaxies in red and star-forming galaxies in blue to emphasize how the star formation properties differ between the two samples. The average mass--size relation from \citet{ELVES-I} and the $1.5\sigma$ line above it are shown in green. The blue and red solid lines show the constant surface brightness line ($\sbeff = 25\ \sbunit$) for galaxies with $g-i=0.4$ and $0.8$ respectively. The two colors are chosen as examples to highlight the dependence of surface brightness cut on galaxy color; for a given stellar mass, a bluer galaxy must have a larger size to be considered as a UDG. Compared with the UPG sample, we find that the UDG sample includes many galaxies that are below the $1.5\sigma$ line and quiescent. Red galaxies dominate the region below the $1.5\sigma$ line because blue galaxies do not satisfy the surface brightness cut in the UDG definition. In contrast, the UPG sample is comprised of many star-forming galaxies at the high-mass end. 

\subsection{Quenched fractions}\label{sec:qfrac}
In this section, we study the quenched fraction as a function of satellite stellar mass. We divide the stellar mass range $6.4 < \log\, M_\star/M_\odot < 9.0$ into eight bins and calculate the fraction of quiescent galaxies for each bin. Both the denominator and numerator are weighted by the importance weight $w_k$ which represents the likelihood of not being a contaminant (see \S\ref{sec:bkg}). We also apply completeness corrections to the denominator and numerator. Not applying the importance weights will make the quenched fraction lower, since the background contaminants are bluer, on average, than UDGs and UPGs. 

\begin{figure}
    \centering
    \includegraphics[width=1\linewidth]{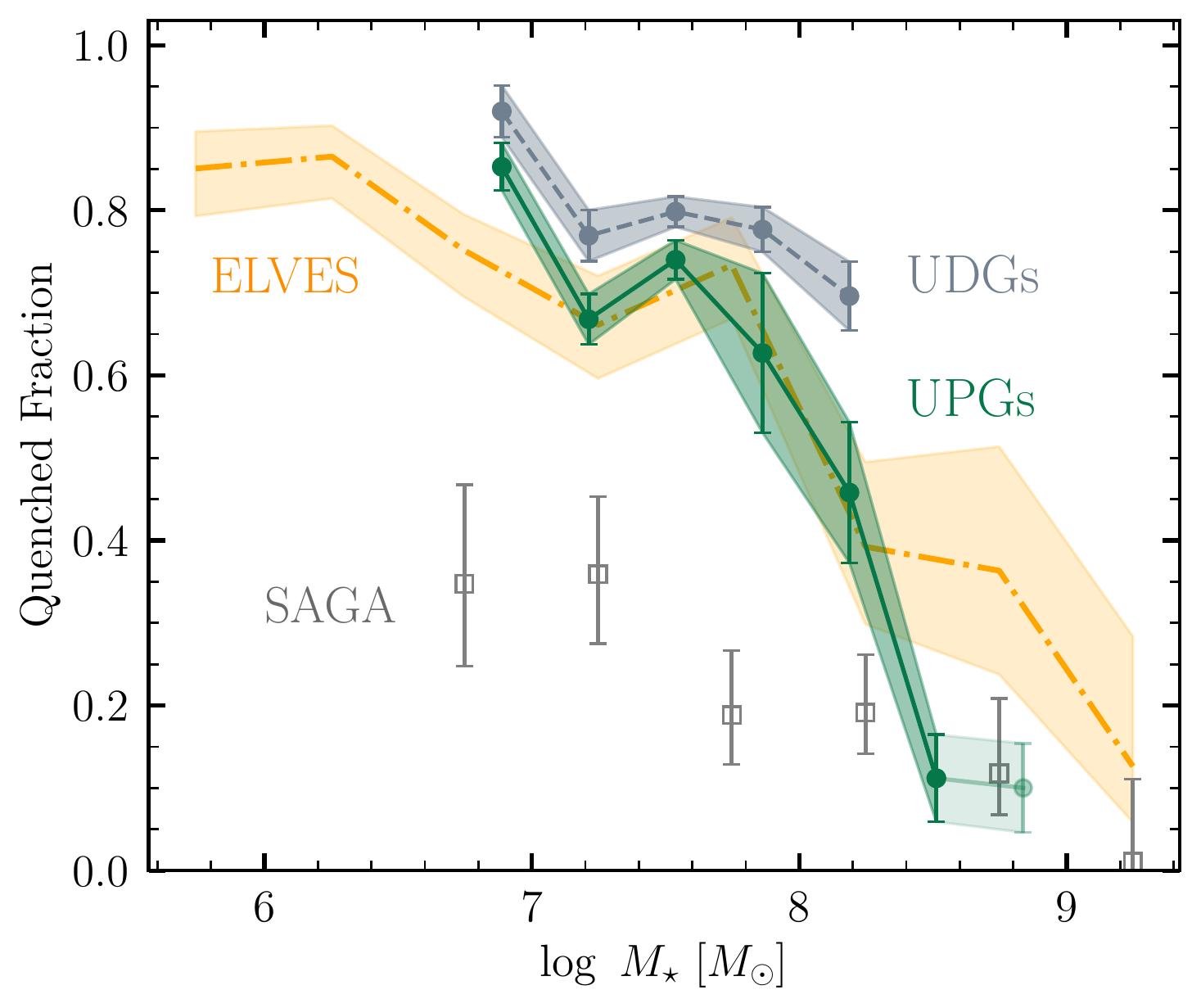}
    \caption{Quenched fractions of UDGs (gray) and UPGs (green) as a function of their stellar masses. The results from ELVES \citep{CarlstenELVES2022} and SAGA \citep{SAGA-II} are also shown for comparison, where quenched fractions are all defined based on the mass-dependent color cut and are corrected for incompleteness following \citet{SAGA-II}. We find that UDGs have a high quenched fraction that changes little over a wide range of stellar mass. On the contrary, the quenched fraction of UPGs is lower and decreases with increasing satellite stellar mass. Given their large sizes, UPGs have a similar quenched fraction to the normal satellites of MW analogs in ELVES.
    }
    \label{fig:qfrac}
\end{figure}

\begin{figure*}
    \vbox{ 
    \centering
    \includegraphics[width=1\linewidth]{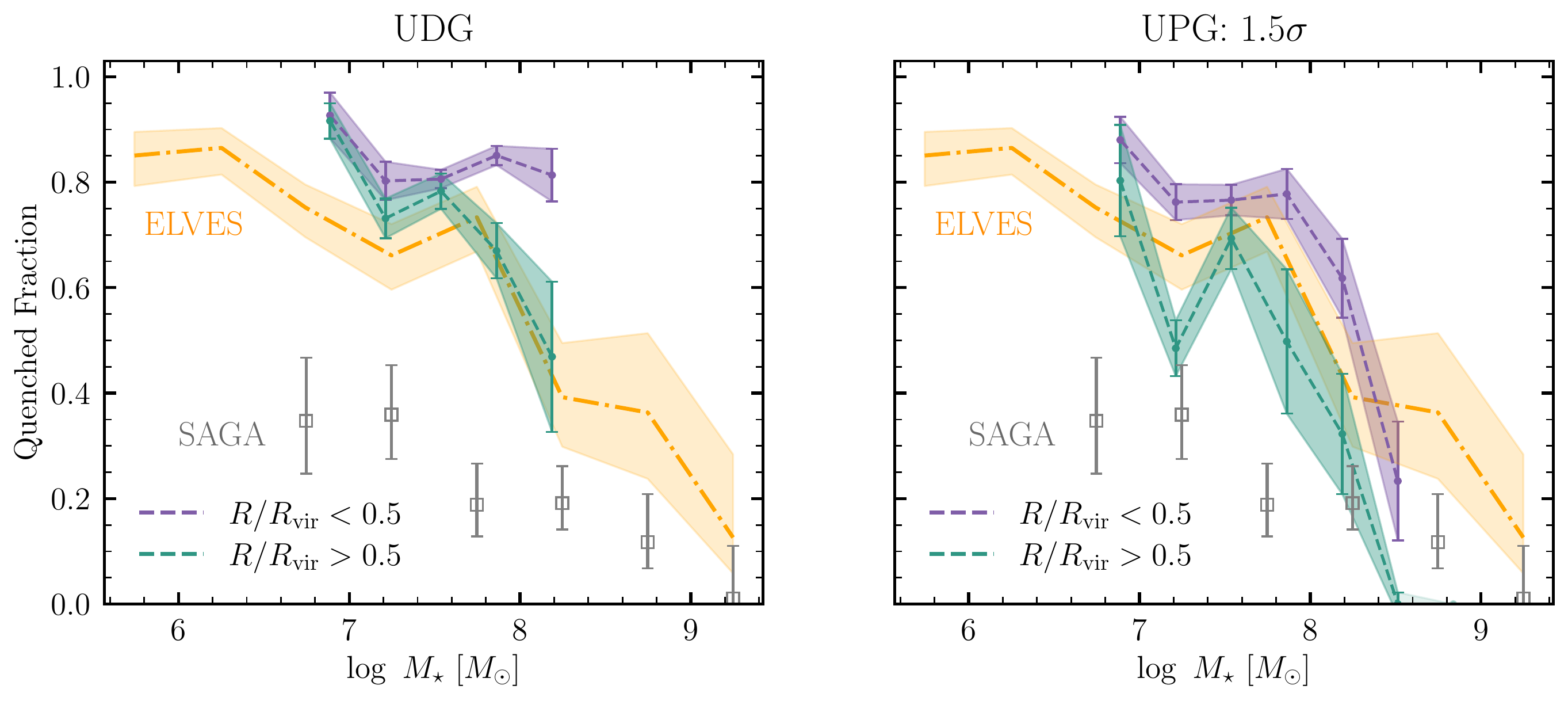}
    \includegraphics[width=1\linewidth]{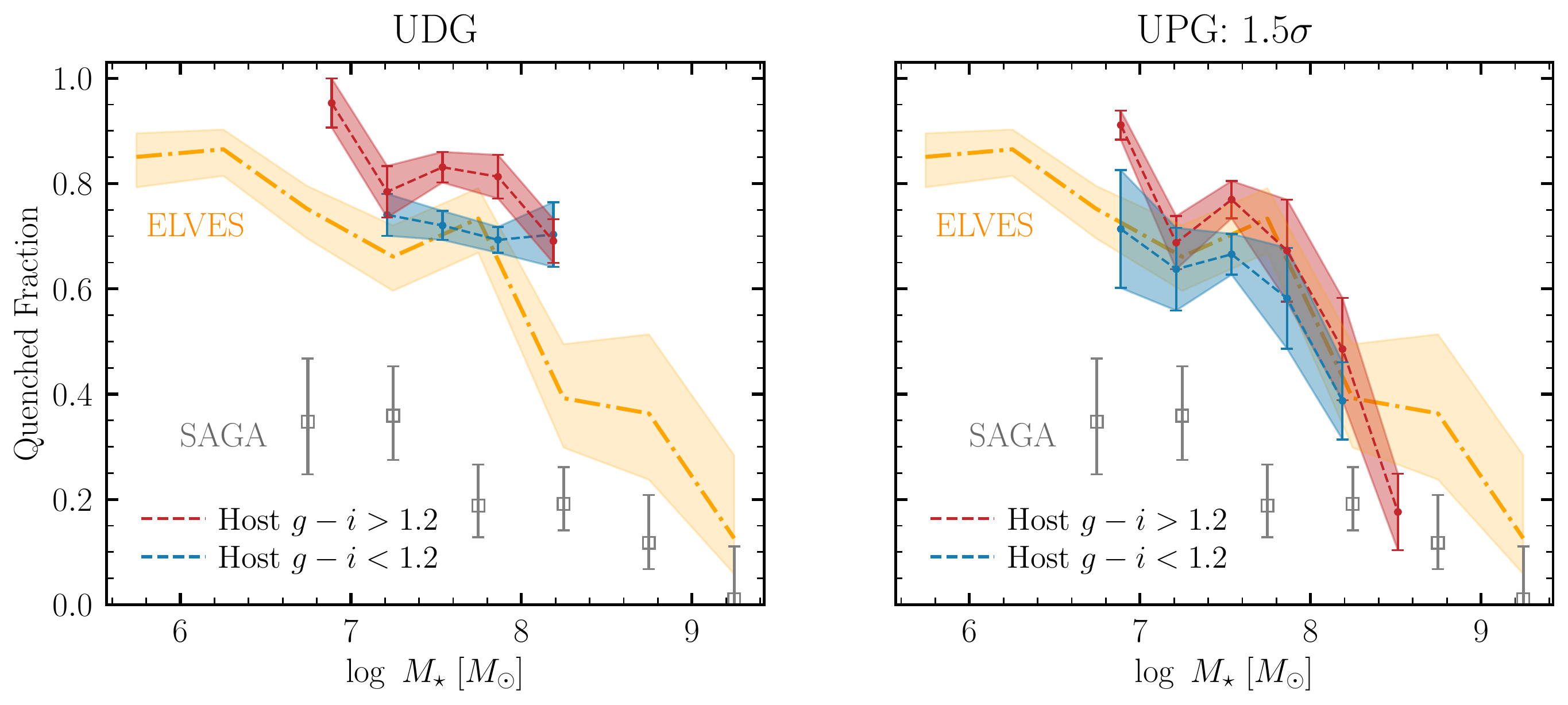}
    }
    \caption{Same as Figure \ref{fig:qfrac}, but the samples are divided into subsets based on the projected distance to the host $R/R_{\rm vir}$ (top panels) and host $g-i$ color (bottom panels), shown as dashed red and blue lines. The quenched fraction is larger for those associated with redder hosts and those that are closer to the host.}
    \label{fig:qfrac_bins}
\end{figure*}

We plot the quenched fractions $f_q$ of the full UDG (gray) and UPG (green) samples in Figure \ref{fig:qfrac}. The error bars correspond to the $1\sigma$ Bernoulli standard error. We plot the UPG quenched fraction above $M_\star > 10^{8.5}\ M_\odot$ with lighter colors to emphasize where we extrapolate the mass--size relation in \citet{ELVES-I} to define UPGs. The quenched fraction of UDGs is high ($f_q \sim 0.8$) and weakly depends on the stellar mass. \citet{Goto2023} also find a high quenched fraction for UDGs among the satellites of MW analogs in the SMUDGes survey. On the contrary, for UPGs, the quenched fraction is lower than that of UDGs and decreases more rapidly with increasing stellar mass. Lower-mass UPGs ($M_\star \sim  10^{7}\ M_\odot$) are mostly quiescent, roughly half of the intermediate-mass UPGs ($M_\star \approx 10^{8}\ M_\odot$) are quenched, and higher-mass UPGs ($M_\star > 10^{8.5}\ M_\odot$) are mostly star-forming. Such distinctions in quenched fractions highlight that UDGs and UPGs select different subsets of the satellite population. In \S\ref{sec:artifact}, we argue that the high quenched fraction of UDGs is merely an artifact of the UDG definition.

We compare the quenched fractions of mass--size outliers (UDGs and UPGs) with those of the satellites of MW analogs in the Local Volume (ELVES; orange line) and the nearby Universe (SAGA; gray squares). Following \citet{CarlstenELVES2022}, the quenched fractions of ELVES and SAGA are calculated based on the same color cut as we use for mass--size outliers\footnote{We note that some galaxies in ELVES and SAGA data only have $g-r$ color because certain surveys might not have $i$-band data. \citet{CarlstenELVES2022} converted $g-r$ color to $g-i$ following their Eqn. (1) in \citet{ELVES-I} that is derived from simple stellar population models.} Thus, the quenched fraction of SAGA in Figure \ref{fig:qfrac} is slightly lower than the values in Fig. 11 of \citet{SAGA-II}, where their quenched fraction is defined based on the detection of H$\alpha$. 
The quenched fractions of SAGA and ELVES are all corrected for incompleteness. We note that the error bars of the SAGA quenched fraction in Figure \ref{fig:qfrac} are smaller than the ones in \citet{SAGA-II}. The tip of the light green error bar in Fig. 11 of \citet{SAGA-II} is derived assuming that all potential satellites without a redshift detection are real quenched satellites. In our case, we adopt the satellite probability for each redshift failure from \citet{SAGA-II} and then apply a color cut to both spectroscopically confirmed and nonconfirmed dwarfs, resulting in a smaller upper bound of the quenched fraction. Overall, the SAGA quenched fraction is lower than that of ELVES and the MW and M31. 
As suggested by \citet{CarlstenELVES2022}, the difference in both luminosity function and quenching could be naturally explained if the SAGA sample is preferentially missing $\sim 1-1.5$ satellites per host, and quiescent galaxies are fainter than star-forming ones at a fixed stellar mass \citep[see also][]{Font2022,Greene2022ELVES}.

From Figure \ref{fig:qfrac}, the quenched fraction of UDGs is higher than that of normal satellites in ELVES and SAGA, especially at the high-mass end ($M_\star \sim 10^{8}\ M_\odot$). However, the quenched fraction of UPGs agrees with ELVES reasonably well at $7 < \log\,M_\star/M_\odot < 8.2$. At the higher-mass end ($\log\,M_\star/M_\odot >8.4$), the UPG quenched fraction agrees with SAGA but is lower than ELVES, with a caveat that the mass--size relation we use might not be valid in this mass range. Overall, it is surprising that UPGs, being mass--size outliers, have a similar quenched fraction as the normal-sized satellites of MW analogs. 

\vspace{1em}
We further divide the sample based on the projected radial distance to the host $R/R_{\rm vir}$ and the host $g-i$ color. As shown in the top panels of Figure \ref{fig:qfrac_bins}, we find that UDGs (UPGs) that are closer to their hosts have higher quenched fractions. The average difference in quenched fractions between the two radial bins is $\sim 0.1$ for UDGs and $\sim 0.2$ for UPGs. This trend agrees with the findings in \citet{Greene2022ELVES} and \citet{Karunakaran2022b} for normal satellites in the ELVES survey. 
From the bottom panels of Figure \ref{fig:qfrac_bins}, we find that UDGs (UPGs) hosted by redder hosts also have a higher quenched fraction. The difference in quenched fractions between the two host color bins is $\sim 0.2$ for UDGs. But the host color has a smaller impact than the distance to the host on the quenched fraction of UPGs. Additionally, we divide the samples based on the host stellar mass and redshift but do not find a significant difference in quenched fractions. 

\begin{figure*}
    \centering
    \includegraphics[width=1\linewidth]{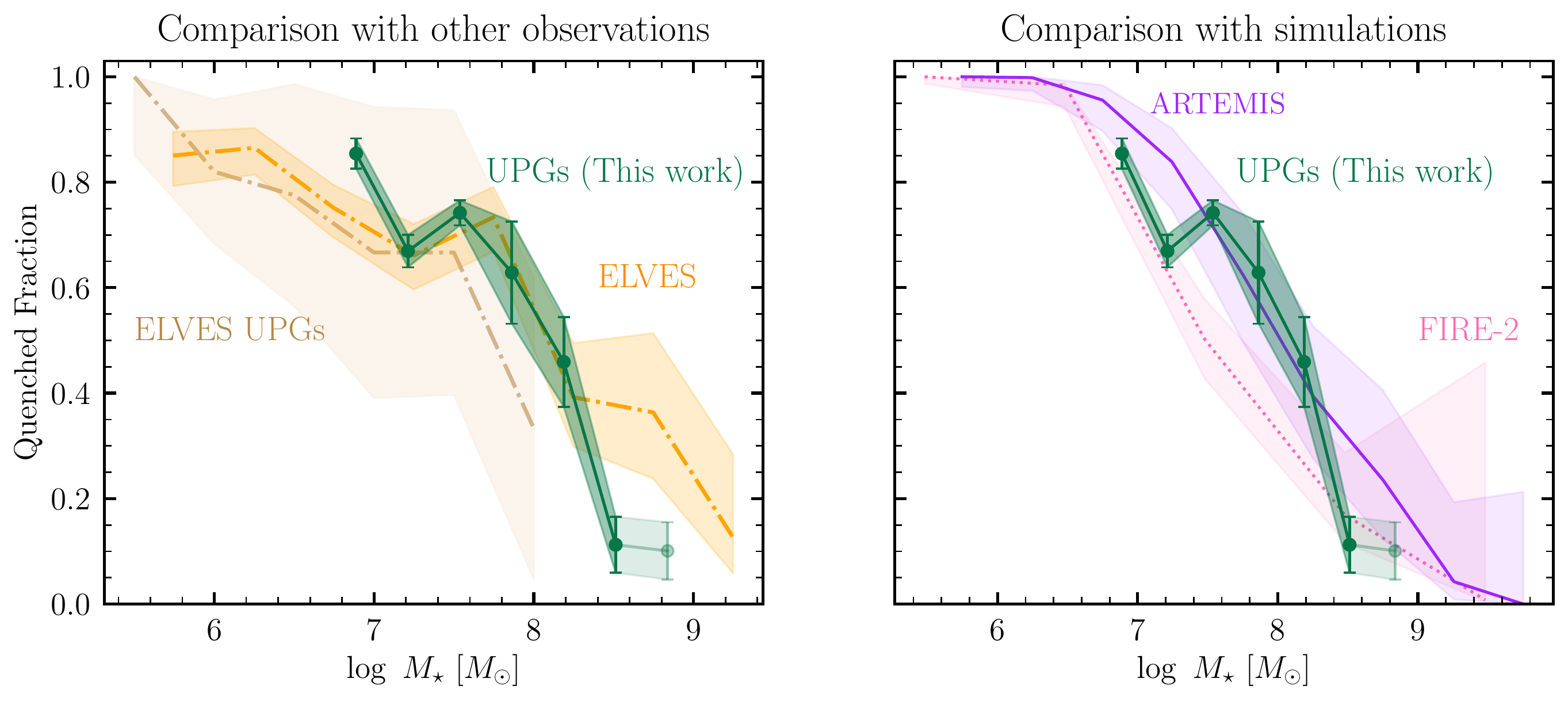}
    \caption{\textit{Left}: quenched fractions of UPGs in this work and in the ELVES survey. We select UPGs from the ELVES sample and calculate the quenched fraction, shown as the brown line. The quenched fraction of UPGs in this work is highlighted in green, while the quenched fraction of the whole ELVES sample is shown in orange. We do not find a significant dependence of the quenched fraction on the size of satellites by comparing UPGs with the bulk of the satellites. \textit{Right}: comparisons with simulations of satellites in MW analogs. The purple line and pink line show the quenched fractions produced in the ARTEMIS \citep{Font2022} and FIRE-2 \citep{Samuel2022} simulations, respectively. The quenched fraction of UPGs agrees quite well with that of normal-sized satellites in simulations. 
    }
    \label{fig:qfrac_elves_sim}
\end{figure*}

\vspace{1em}
As a sanity check, we select UPGs in the ELVES sample and calculate the quenched fraction, shown as the brown line in the left panel of Figure \ref{fig:qfrac_elves_sim}. Out of the 313 ELVES satellites that have secure distances around hosts that are mapped out to $R>200$ kpc, there are 27 UPGs \citep[see][]{Greene2022ELVES}. Despite the number of UPGs in ELVES being small, the quenched fraction of ELVES UPGs agrees with that of our UPG sample (green line) quite well. We also do not find a significant difference between the quenched fraction of the whole ELVES sample from that of ELVES UPGs. 

We compare the quenched fraction of UPGs with the simulation results from ARTEMIS \citep[purple lines in Figure \ref{fig:qfrac_elves_sim};][]{Font2022} and FIRE-2 \citep[pink lines in Figure \ref{fig:qfrac_elves_sim};][]{Samuel2022}, which focus on studying normal-sized satellites of MW analogs. The quenched galaxies in these simulations are defined as those with no instantaneous (for ARTEMIS) or recent ($<200$ Myr for FIRE-2) star formation.\footnote{\citet{Font2022} found a similar quenched fraction for ARTEMIS when using the same color cut as we do.} We select ARTEMIS and FIRE-2 among other simulations because they roughly cover the highest and lowest quenched fraction for MW satellites produced in simulations (see Fig. 13 in \citealt{Samuel2022} for a compilation). Since none of these data sets are designed to study mass--size outliers, their quenched fractions are presumably dominated by normal-sized satellites. As shown in the right panel of Figure \ref{fig:qfrac_elves_sim}, our UPG quenched fraction agrees best with ARTEMIS but is also consistent with the FIRE-2 result. We discuss the implications of these results in \S\ref{sec:puff}.

\section{Discussion}\label{sec:discussion}

\subsection{The High Quenched Fraction of UDGs Is an Artifact}\label{sec:artifact}

In \citetalias{Li2022}, we propose the concept of ultra-puffy galaxies (UPGs) as a means to robustly study mass--size outliers. Compared with the UDG selection, UPGs are not selected via a constant cut in half-light radius. The UPG selection considered the fact that the average sizes of satellite galaxies increase with stellar mass following a mass--size relation. There are many UDGs that are not outliers with respect to the mass--size relation. Thus, the UPGs better represent the tail of the satellite size distribution. Moreover, for a given stellar mass and size, blue galaxies have brighter surface brightness than red galaxies because red galaxies have a higher mass-to-light ratio. As a consequence, blue galaxies will be preferentially excluded if one applies a hard cut on surface brightness, as in the definition of UDG. As seen in Figure \ref{fig:mass_size}, the UDG sample includes a number of galaxies below the $1.5\sigma$ line, and these galaxies are mostly red and quenched, therefore giving rise to a high quenched fraction in Figure \ref{fig:qfrac}. 
This high quenched fraction of UDGs compared to normal-sized dwarfs is just an artifact of the UDG definition. The hard size cut and surface brightness cut make it difficult to directly compare the red and blue galaxy populations and hamper the study of diffuse galaxies. The cut at $1.5\sigma$ above the average mass--size relation does not discriminate blue or red galaxies but just selects size outliers at a given stellar mass. The UPGs are physically motivated and naturally represent the large-size tail of satellite galaxies in MW analogs. 

We note that \citet{Lim2020} defined UDGs in the Virgo cluster as the outliers of the scaling relations of Virgo satellites. However, in clusters, the difference between such a definition and the fiducial UDG definition in \citet{vanDokkum2015} becomes minor because nearly all UDGs are red and quenched. Nevertheless, we advocate studying outliers of scaling relations, including UPGs, which provide an unbiased perspective on the formation and evolution of diffuse dwarf galaxies. 

\subsection{Formation and Quenching of mass--size Outliers}\label{sec:puff}

In this paper, we study the mass--size outliers associated with MW analogs in the nearby Universe. We calculate the quenched fraction of UPGs in \citetalias{Li2022} and compare it with the ELVES survey, the SAGA survey, and numerical simulations. One key finding is that UPGs have a similar quenched fraction as those normal-sized satellites of MW analogs in the Local Volume (ELVES; Figure \ref{fig:qfrac}) and numerical simulations (ARTEMIS and FIRE-2; Figure \ref{fig:qfrac_elves_sim}). We also check the quenched fraction of UPGs among ELVES satellites and find similar results (Figure \ref{fig:qfrac_elves_sim}). In addition, the radial distribution of mass--size outliers can be well described by an NFW (or Einasto) profile at $R > 0.2\ R_{\rm vir}$. 
Thus, it is intriguing that despite being selected as outliers in size, UPGs are acting like ``normal'' satellites in terms of quenched fraction and radial distribution. Furthermore, \citet{ELVES-I} found that the mass--size relation of satellites in MW analogs does not depend on the morphology or color of the satellites. All of these pieces of evidence suggest that, for satellites in MW analogs, quenching is not tied to being a mass--size outlier. Quenching and morphological transformation might involve very mild size evolution. 

We are left with the question of what physical processes are responsible for puffing up satellites and quenching them such that quenching and size growth happen separately. If the mass--size outliers were normal-sized before falling into the group, then tidal interactions are believed to puff them up. \citet{Jiang2019} found that in simulations, the UDGs in groups are accreted as either UDGs from the field or normal dwarfs but tidally heated near the orbital pericenter. \citet{Tremmel2020} studied simulated UDGs in the cluster environment and found that UDGs with higher mass ($M_\star > 10^{8}\ M_\odot$) are puffed up suddenly due to tidal heating at the pericenter, but UDGs with lower mass ($M_\star \sim 10^{7.5}\ M_\odot$) are gradually puffed up due to adiabatic expansion as a response to the mass loss from tidal and ram pressure stripping. In this scenario, normal-sized star-forming dwarf galaxies are accreted from the field, then consequently puffed up by strong tidal interactions. If so, we would expect a higher quenched fraction for mass--size outliers, since they must have undergone violent ram pressure stripping near the pericenter or lost gas due to tidal stripping. This contradicts our results in Figure \ref{fig:qfrac}.

On the other hand, the mass--size outliers can be puffed up when they are still in the field. For example, bursty stellar feedback can cause the stellar component of the dwarf galaxy to expand \citep[e.g.,][]{DiCintio2017,Chan2018,Carleton2019,Jiang2019,Martin2019}. Analytical models also support a scenario where UDGs originate from a population of dwarf galaxies residing in halos with higher spin \citep{Dalcanton1997,Amorisco2016,Rong2017,Liao2019}. \citet{Wright2021} showed that field UDGs can also be formed from major mergers that cause star formation to migrate outward. If mass--size outliers are already puffed up prior to infall, in order to achieve a similar quenched fraction as normal satellites, they must remain star-forming after being puffed up such that they can be quenched together with normal satellites. \citet{Wright2021} showed that an early merger does not significantly change the total SFR. \citet{Samuel2022} showed that preprocessing can boost the quenched fraction by 20\% at $M_\star \approx 10^8\ M_\odot$ and merely change the quenched fraction for lower-mass satellites. Although it is not clear whether the high halo spin is the cause or the consequence of UPG formation, it is probable that a higher halo spin does not quench a UPG. If any of these mechanisms can only increase the size to make UPGs but not quench them when they are still in the field, it provides a plausible way to produce the observed trend in Figures \ref{fig:qfrac} and \ref{fig:qfrac_elves_sim}.

Mass--size outliers and normal satellites might be quenched together after falling into the group. It is not obviously clear whether they are quenched mainly by tidal or ram pressure stripping. In the tidal stripping scenario, the dark matter needs to be removed before a significant fraction of stars and/or gas can be stripped by tidal forces, indicating a longer quenching timescale. \citet{Jiang2019} argued that if tidal stripping dominates the quenching of UDGs, the satellites closer to the host will have a smaller stellar mass and effective radius, since stars are also stripped. They found no such trend in simulations of group environments. In \S\ref{sec:size_distr}, we split the UDG/UPG samples into two radial distance bins but do not find a significant change in size or stellar mass distribution as the mass--size outliers get closer to the host. This suggests that tidal stripping does not dominate the quenching of UDGs and UPGs in MW analogs. 

Many studies support the idea that ram pressure stripping is the dominant quenching mechanism for normal-sized satellites with $M_\star < 10^{8.0}\ M_\odot$ \citep[e.g.,][]{Tonnesen2009,Jaffe2015,Simpson2018,Akins2021,WangJ2021,Samuel2022}. In the Auriga simulations, for instance, \citet{Simpson2018} showed that the quenched fraction of satellites $M_\star < 10^8\ M_\odot$ have a strong dependence on the distance to the hosts, and they considered ram pressure stripping as the main quenching mechanism. Using the FIRE-2 simulation, \citet{Samuel2022} also showed that the quenched fraction increases as the distance to the host decreases, and hosts with higher CGM mass have more quenched satellites. Such trends are also found in observations \citep[e.g.,][]{Greene2022ELVES,Karunakaran2022c}. In Figure \ref{fig:qfrac}, by comparing UDGs and UPGs at two radial bins, we find a very similar trend that satellites that are closer to the host are more quiescent. Although we are not able to derive a quenching timescale to better compare with simulations, we consider our results to be consistent with the ram pressure stripping scenario. 

In Figure \ref{fig:qfrac_bins}, we also find that UDGs and UPGs associated with redder hosts are more quiescent than those in bluer hosts, although the trend is less significant for UPGs. This trend agrees with ``galaxy conformity,'' which refers to the excess of early-type satellites in the vicinity of early-type central galaxies compared to satellites around late-type centrals \citep{Weinmann2006}. \citet{Wang2012} argued that galaxy conformity arises because quiescent centrals occupy more massive halos than star-forming centrals. More massive halos have more hot gas and stronger tidal fields, leading to more efficient quenching. If quiescent centrals are in older halos, their satellites tend to be accreted earlier and thus have more time to be quenched. Based on such models, the galaxy conformity signals from UDGs and UPGs make sense if they are puffed up before the infall and quenched together with normal satellites due to ram pressure stripping. 

The efficiency of ram pressure stripping also depends on the stellar mass density, which provides the restoring force and the gas mass density. The diffuse stellar component might make mass--size outliers more vulnerable to ram pressure and tidal stripping. However, \citet{Kado-Fong2022UDG} found in observations that although UDGs have a more diffuse stellar component, they harbor the same amount of \ion{H}{1} gas as normal dwarfs with similar stellar sizes. Equivalently, for a given stellar mass, UDGs (UPGs) have higher \ion{H}{1} mass than normal dwarfs. Thus, even if UDGs and UPGs are more vulnerable to ram pressure stripping, they have more gas to be stripped. It is probable that the net effect of low stellar mass density and high \ion{H}{1} mass is that UPGs have a similar quenched fraction as normal satellites. Of course, the origin and evolution of the mass--size outliers in MW analogs could also be a mixture of many scenarios. Using our observations, it is hard to sort out the detailed mechanisms for quenching and size growth. More detailed observations and simulations are needed to solve this intriguing puzzle.


\section{Summary}\label{sec:summary}
In this work, we present a statistical analysis of mass--size outliers among the satellite galaxies of MW analogs at $0.01 < z < 0.04$ based on the sample from \citetalias{Li2022}. Besides the ultra-diffuse galaxies (UDGs), we study ``ultra-puffy galaxies'' (UPGs), which are defined to lie $1.5\sigma$ above the average mass--size relation and better represent the large-size tail of the dwarf galaxy population. We derive the size distribution (\S\ref{sec:size_distr}), radial distribution (\S\ref{sec:radial_distr}), and quenched fractions (\S\ref{sec:quench}) of mass--size outliers (UDGs and UPGs) and discuss implications on their formation and evolution (\S\ref{sec:discussion}). We summarize our main findings and prospects as follows.

\begin{enumerate}
    \item The size distribution of UDGs follows a power law $\mathrm{d} n / \mathrm{d}\log r_e \propto r_e^{-1.45 \pm 0.20}$ which is shallower than the UDG size distribution in larger groups and clusters \citep[e.g.,][]{vdBurg2016,vdBurg2017}. Since the UPG sample includes fewer small satellites, its size distribution is suppressed at the smaller-size end (Figure \ref{fig:size_distribution}). We do not find significant dependence of size distribution on the host stellar mass or the distance to the host. 
    
    \item The radial distributions of UDGs and UPGs can be described by projected NFW and Einasto profiles (Figure \ref{fig:radial_distribution}). UDGs and UPGs follow similar radial distributions. We find that the best-fit NFW profiles have concentrations similar to that of an MW-mass halo. The best-fit Einasto profiles are less concentrated than MW-mass halos. 
    
    \item The UDGs have a very high quenched fraction ($\sim 70\%$) that remains roughly constant over 1 dex in satellite stellar mass (Figure \ref{fig:qfrac}). However, for UPGs, the quenched fraction shows a strong dependence on satellite stellar mass; 80\% of UPGs are quiescent at the lower-mass end $M_\star = 10^{6.8}\ M_\odot$, but the fraction drops to 20\% at $M_\star = 10^{8.5}\ M_\odot$. Almost half of UPGs are quenched at $M_\star \approx 10^{8}\ M_\odot$.  
    The quenched fractions of both UDGs and UPGs increase as satellites get closer to the hosts, agreeing with the trends of normal-sized satellites \citep{Greene2022ELVES,Karunakaran2022c}. Redder hosts also harbor more quenched mass--size outliers, showing evidence for ``galaxy conformity.'' 
    
    \item Surprisingly, although UPGs have a much larger size than ``normal'' satellites of similar stellar mass, they show a similar quenched fraction as the normal-sized satellites in MW analogs (Figure \ref{fig:qfrac_elves_sim}). The UPG quenched fraction is consistent with the results from both observations (ELVES) and simulations (ARTEMIS, FIRE-2) for normal-sized satellites. 
    Such a null trend of the quenched fraction on the size of satellites indicates that the quenching and size evolution of mass--size outliers happen separately in MW-like environments. It is plausible that these mass--size outliers are puffed up in the field before accretion and quenched together with normal satellites (\S\ref{sec:puff}). 
    
    \item We argue that the high and constant quenched fraction for UDGs is merely an artifact of the UDG definition, where the constant surface brightness cut preferentially selects more red galaxies (Figure \ref{fig:mass_size}). We demonstrate that the more physically motivated UPG selection does not introduce artifacts in the mass--size distribution and better represents the large-size tail of the dwarf galaxy population. 
\end{enumerate}

\section*{Acknowledgment}
We thank the anonymous reviewer for useful comments that make the manuscript much clearer. J.L. is grateful for the discussions with Meng Gu, Frank van den Bosch, and Sihao Cheng. J.L. also thanks David Tao and his \textit{Soul Power} live album. J.E.G gratefully acknowledges support from NSF grant AST-1007052. S.D. is supported by NASA through Hubble Fellowship grant HST-HF2-51454.001-A awarded by the Space Telescope Science Institute, which is operated by the Association of Universities for Research in Astronomy, Incorporated, under NASA contract NAS5-26555.

The Hyper Suprime-Cam (HSC) collaboration includes the astronomical communities of Japan and Taiwan and Princeton University. The HSC instrumentation and software were developed by the National Astronomical Observatory of Japan (NAOJ), Kavli Institute for the Physics and Mathematics of the Universe (Kavli IPMU), University of Tokyo, High Energy Accelerator Research Organization (KEK), Academia Sinica Institute for Astronomy and Astrophysics in Taiwan (ASIAA), and Princeton University.  
Funding was contributed by the FIRST program from the Japanese Cabinet Office, Ministry of Education, Culture, Sports, Science and Technology (MEXT), Japan Society for the Promotion of Science (JSPS), Japan Science and Technology Agency (JST), Toray Science Foundation, NAOJ, Kavli IPMU, KEK, ASIAA, and Princeton University. The authors are pleased to acknowledge that the work reported on in this paper was substantially performed using the Princeton Research Computing resources at Princeton University, which is a consortium of groups led by the Princeton Institute for Computational Science and Engineering (PICSciE) and the Office of Information Technology's Research Computing.

\vspace{1em}
\software{\href{http://www.numpy.org}{\code{NumPy}} \citep{Numpy},
          \href{https://www.astropy.org/}{\code{Astropy}} \citep{astropy}, \href{https://www.scipy.org}{\code{SciPy}} \citep{scipy}, \href{https://matplotlib.org}{\code{Matplotlib}} \citep{matplotlib},
          \href{https://statmorph.readthedocs.io/en/latest/}{\code{statmorph}} \citep{statmorph},
          \href{https://halotools.readthedocs.io/en/latest}{\code{Halotools}} \citep{Hearin2017},
          \href{https://bdiemer.bitbucket.io/colossus/index.html}{\code{Colossus}} \citep{Diemer2018},
          \href{https://pmelchior.github.io/scarlet/}{\code{scarlet}} \citep{Melchior2018}, \href{https://github.com/dr-guangtou/unagi}{\code{unagi}}.
          }

\bibliography{citation}{}
\bibliographystyle{aasjournal}

\end{CJK*}
\end{document}